# Transient Stability Analysis and Fault Clearing Angle Estimation of VSG Based on Domain of Attraction Estimated by Trajectory Reversing Method

Jiayue Lyu, Tianzhi Fang, *Member, IEEE*, Zhiheng Lin, *Member, IEEE,* Jingxue Han, and Yantao Zhu

*Abstract*-- The virtual synchronous generator (VSG), with the analogous nonlinear power-angle relationship to the synchronous generator (SG), has attracted much attention as a promising solution for converter-based power systems. In this paper, a large signal model of the grid-connected VSG is first established. The trajectory reversing method (TRM) is then introduced to estimate the domain of attraction (DOA) of VSG. Subsequently, the transient instability mechanism is revealed in detail based on the estimated DOA boundary. The impacts of system parameters on the DOA range are further investigated. It is found that loss of synchronization (LOS) occurs if the system trajectory lies outside the post-fault DOA range. In scenarios where no equilibrium points exist after a grid fault, system stability can be reestablished only when the fault clearing angle (FCA) does not exceed the critical clearing angle (CCA). Finally, the CCA derived from the DOA and that from the conventional equal area criteria (EAC) are compared. The results show that CCA obtained by our solution has a higher accuracy. Time-domain simulations are performed to verify the effectiveness of the proposed transient stability analysis method of grid-connected VSG.

*Index Terms*-- Transient stability, virtual synchronous generator (VSG), domain of attraction (DOA), trajectory reversing method (TRM), inverter.

## I. INTRODUCTION

THE increasing penetration of power electronic converters has led to remarkable changes in the conventional power grid structure, which is primarily based on synchronous generators (SGs). These converters, characterized by low inertia and underdamping, present a notable impact on the stability of a high proportion of distributed energy resources (DERs) systems [1]. As the ratio of converter-based DERs increases, the inertia of the power grid is gradually decreasing, leading to a deterioration of the frequency stability during grid disturbances [2]. To provide the inertia and damping support for the power grid, virtual synchronous generator (VSG) is proposed to emulate the mechanical damping and rotating inertia characteristics of SGs. Moreover, VSGs are capable of providing frequency regulation and voltage support to a power system, and operating in both grid-connected and standalone mode [3]-[4]. Based on these advantages, the application of VSG is gradually increasing. However, after the occurrence of a large signal disturbance, the oscillation of output power and frequency can also be found in VSGs, similar to SGs [5]. Therefore, it is essential to investigate the transient instability mechanism of VSG, identify the predominant parameters affecting the system's stability, and analyze how the parameters affect the system's stability.

In order to study the transient process of the VSG after a large signal disturbance and evaluate its stability, several analytical methods have been employed separately. Because of the similarity between the dynamic equation of VSG and the swing equation of SG, the equal area criteria (EAC), extensively utilized in SG stability analysis, has been adapted for the transient stability analysis of VSG [8]. The loss of synchronization (LOS) can be determined by whether the acceleration areas exceed the deceleration areas. However, the EAC method is only valid under the assumption that the system damping coefficient is zero, which inevitably introduces conservatism to the system [9]. Therefore, the accuracy of its application to VSGs is compromised by the presence of a large damping coefficient [7],[9]. An extended EAC method considering the damping coefficient to evaluate the transient stability of VSG is introduced in reference [29], which refines the accuracy of the stability assessment. However, the calculation of acceleration and deceleration areas becomes more complicated.

The phase portrait method stands as another analytical tool, effectively addressing the limitations inherent in the EAC by taking the damping coefficient into account [7]. In addition, the phase portrait method does not need to compare the size of the acceleration area with the deceleration area. For second-order nonlinear differential equations which cannot be solved directly, the phase portrait method offers an intuitive visualization of the system trajectory in the phase plane for a given initial condition [11]. LOS determines whether the phase trajectory converges to a new equilibrium point. Whereas, although the phase portrait method can determine the transient instability with higher accuracy, it lacks the ability to explain the mechanism of system instability.

To deal with the inadequacy of phase portrait method, the concept of the domain of attraction (DOA) is introduced. DOA is the set of all initial conditions that asymptotically converge towards an equilibrium point, typically depicted as a region in two-dimensional space [19]. The exact DOA region is conducive to the stability assessment. Thus, in recent years, numerous research interests have been attracted to the DOA region estimation. The energy function and Lyapunov direct method have been the most commonly adopted solutions

J. Lyu, T. Fang, J. Han, Y. Zhu are with the Center for More-Electrical-Aircraft Power System, College of Automation Engineering, Nanjing University of Aeronautics and Astronautics, Nanjing 211106, China (e-mail: {stardust; fangtianzhi; 2403075 and 022nuaazyt20417}@nuaa.edu.cn).

Z. Lin is with the Department of Electrical and Computer Engineering, University of Alberta, Edmonton, Alberta T6G 2R3, Canada (e-mail: zhiheng.lin@ualberta.ca).

[8],[13]-[18]. Similar to EAC, the energy function method ignores the nonlinear damping term, which may render the stability analysis results either overly conservative or overly aggressive [9]. Lyapunov direct method excels in providing a quantitative assessment of nonlinear system stability, especially in high dimension systems [8]. Nonetheless, the construction of an appropriate Lyapunov function remains a formidable challenge due to the absence of a generic and systematic construction standard [17]. The estimated DOA region highly depends on the specific form of the constructed Lyapunov function, with an inherent degree of conservatism [13], [14], [17].

Trajectory reversing method (TRM) is an effective technique to analyze dynamic systems by tracing the system trajectory backward. This method is capable of yielding an approximate DOA estimation that is remarkably precise in second-order systems [15],[20]-[21]. Several works have been dedicated to estimating the DOA boundary for phase-locked loop (PLL) based voltage source converters (VSCs) [22]-[24]. However, the application of similar methodologies in the context of grid-connected VSGs remains relatively scarce. In light of the limitations inherent in the aforementioned transient stability analysis methods, this paper employs TRM to estimate the boundary of DOA. Furthermore, this paper conducts an in-depth discussion on the transient stability criterion and the underlying instability mechanisms. The main contributions of this paper are summarized as follows:

1) Firstly, the precise DOA boundary of VSG is acquired by means of TRM estimation. Based on the DOA region estimated by TRM and the phase trajectories of the system during large signal perturbations, the transient instability mechanism of VSG is analyzed in detail.

2) Secondly, theoretical analysis is further conducted on the variation of the DOA boundary as system parameters fluctuate, which is beneficial for providing a profound insight into the transient instability conditions.

3) Thirdly, the critical clearing angle (CCA) derived from DOA is compared against that obtained by conventional EAC. The comparative results reveal that CCA obtained through our solution exhibits greater precision.

The structure of the subsequent sections of this paper is outlined as follows. In Section II, a comprehensive derivation of the dynamic representation of the power angle is presented based on the system diagram. Then, types of transient stability problems are analyzed and TRM is then introduced to estimate the DOA boundary of VSG. An in-depth exploration of the impact of various system dynamic equation parameters on the DOA boundary is also provided. Section III focuses on the analysis of transient stability for two categories of transient stability issues under diverse large disturbances. CCA, as determined through the DOA, is compared with that obtained by conventional EAC to highlight the comparative efficacy. In Section IV, time-domain simulations are given to verify the theoretical analysis presented above. Finally, conclusions are summarized in Section V.

## II. LARGE-SIGNAL MODELING OF VSG

### A. System Description

Fig.1(a) illustrates the configuration of a grid-connected VSG, where $L_f$ and $C_f$ constitute the output filter of the VSG, while $i_L$ and $i_g$ denote the output current of the VSG and the current injected into the grid, respectively. $V_{PCC}$ is the amplitude of the voltage at the point of common coupling (PCC). $R_g$ and $X_g$ denote grid resistance and reactance, where $X_g = \omega_0 L_g$, $\omega_0$ denotes the rated angular speed. $V_{dc}$ is the input dc voltage and it can be assumed constant when analyzing the interaction between the grid-connected VSG and the power grid on account of the addition of the energy storage unit and renewable energy sources into the VSG [25].

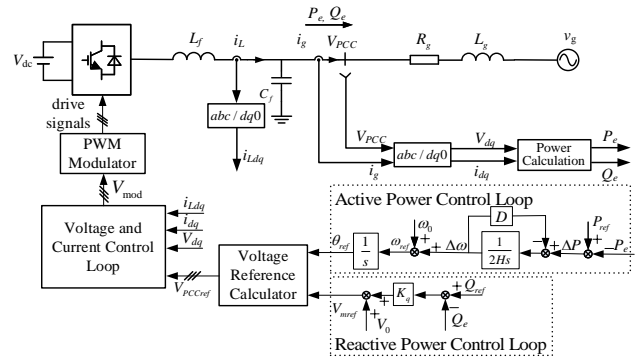

(a) Single-line diagram of VSG

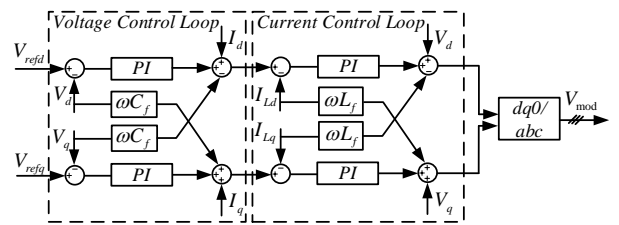

(b) Voltage and current control structure

Fig. 1. System topology of a grid-connected VSG.

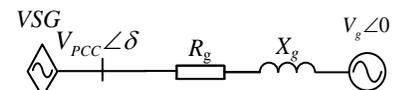

Fig. 2. Equivalent circuit of the grid-connected VSG.

As depicted in Fig.1(b), the outer voltage control loop is used to ensure the PCC voltage $V_{PCC}$ tracks its reference $V_{PCCref}$ and the inner current loop is adopted to achieve current limitation naturally [26]. The transient stability of VSG is mainly determined by the outer power control loop because the outer power control loops are engineered with a significantly lower bandwidth compared to the inner voltage-current control loop. The inner voltage and current control loop can be generally assumed as a unit gain with ideal tracking performance during the transient stability analysis. Thus, we can assume that $V_{PCC} = V_{PCCref}$. The phase reference $\theta_{ref}$ is generated by the active power control loop and the reactive power-voltage(Q-V) droop control is used to generate the voltage magnitude reference $V_{mref}$.

Based on the active power control loop, we have

$$\begin{cases} \Delta P = P_{ref} - P_e \\ \Delta \omega = \dfrac{1}{2Hs}(\Delta P - D \cdot \Delta \omega) \\ \omega_{ref} = \omega_0 + \Delta \omega \\ \theta_{ref} = \dfrac{1}{s} \cdot \omega_{ref} \end{cases} \quad (1)$$

where $P_{ref}$ and $P_e$ denote the active power reference and active power output of the VSG, respectively; $H$ is the inertia constant, and $D$ is the damping coefficient; $\omega_{ref}$ is the angular frequency reference; $\Delta \omega$ denotes the angular frequency deviation.

Defining the angle difference between $\theta_{PCC}$ and $\theta_g$ as power angle $\delta$, which yields

$$\delta = \theta_{PCC} - \theta_g \quad (2)$$

Considering $\theta_{ref} = \theta_{PCC}$ and substituting (2) into (1), the dynamic equation of the VSG can be expressed as:

$$2H\ddot{\delta} = P_{ref} - P_e - D\dot{\delta} \quad (3)$$

According to the reactive droop control diagram, the voltage magnitude reference $V_{mref}$ is given by

$$V_{mref} = V_0 + K_q(Q_{ref} - Q_e) \quad (4)$$

where $V_o$ represents the nominal voltage magnitude. $Q_{ref}$ and $Q_e$ are the reactive power reference and reactive power output of the VSG, respectively. $K_q$ denotes the Q-V droop coefficient.

Fig.2 illustrates the equivalent circuit of the grid-connected VSG where the VSG is modeled as a controlled voltage source. According to Fig.2, the output active and reactive power of the VSG can be calculated as

$$P_e = \frac{3}{2} \cdot \frac{V_{PCC}}{R_g^2 + X_g^2}\left[R_g(V_{PCC} - V_g \cos\delta) + X_g V_g \sin\delta\right] \quad (5)$$

$$Q_e = \frac{3}{2} \cdot \frac{V_{PCC}}{R_g^2 + X_g^2}\left[X_g(V_{PCC} - V_g \cos\delta) - R_g V_g \sin\delta\right] \quad (6)$$

By substituting (6) into (4), we can obtain

$$V_{PCC} = V_0 + K_q\left(Q_{ref} - \frac{3}{2} \cdot \frac{V_{PCC}}{R_g^2 + X_g^2}\left[X_g(V_{PCC} - V_g \cos\delta) - R_g V_g \sin\delta\right]\right) \quad (7)$$

Considering $V_{PCC}=V_{mref}$, we can deduce the relationship between $V_{PCC}$ and $\delta$, which is expressed as (8) shown at the bottom of this page. Besides, by substituting (5) and (8) into (3), the second-order nonlinear differential equation of the grid-connected VSG can be obtained, which is presented by (9) at the bottom of this page.

*B. Types of Transient Stability Problems*

In order to investigate the transient response and instability mechanisms of VSG during transient processes, it is imperative to categorize the possible fault scenarios encountered under voltage drop faults. According to (4) and the parameters given in Table I, the $P_e$-$\delta$ curves of the pre-fault VSG can be depicted in Fig.3 with the dashed red line. Meanwhile, the solid black line represents the post-fault $P_e$ curve. As depicted in Fig.3, transient stability issues can generally be categorized into three distinct types, contingent upon the presence or absence of equilibrium points during the large disturbances.

***Type-I and II*** (With equilibrium points): Fig.3 (a) and (b) show the scenarios where equilibrium points exist after grid fault occurs. Point *b* and *d* represent the stable equilibrium point (SEP) and unstable equilibrium point (UEP), respectively. The system is initially operated at SEP *a* where active power is balanced. When the fault occurs, the operating point of the system moves from the point *a* to the point *b* instantaneously. Given that $P_{ref} > P_e$ at point *b*, the output frequency of the VSG starts to increase until it reaches SEP *c*. During this period, an acceleration area $S_{acc}$ can be observed. When the VSG crossover the SEP *c*, the output frequency of the VSG begins to decrease due to $P_{ref} < P_e$, and a deceleration area $S_{dec}$ emerges accordingly. According to EAC, if $S_{acc} < S_{decmax}$, the VSG can synchronize to the grid frequency before UEP *d*, and thus, the VSG will maintain stable during the fault. Conversely, if $S_{acc} > S_{decmax}$, the output frequency and $\delta$ will further increase at this time. Consequently, the VSG will lose synchronization.

$$V_{PCC} = \frac{1.5K_q V_g(X_g \cos\delta + R_g \sin\delta) - (R_g^2 + X_g^2)}{3K_q X_g} + \frac{\sqrt{\left(1.5K_q V_g(X_g \cos\delta + R_g \sin\delta) - (R_g^2 + X_g^2)\right)^2 + 6K_q X_g(V_0 + K_q Q_{ref}) \cdot (R_g^2 + X_g^2)}}{3K_q X_g} \quad (8)$$

$$2H\ddot{\delta} = -D\dot{\delta} + P_{ref} - \frac{1.5K_q X_g V_g \cos\delta + 1.5K_q R_g V_g \sin\delta - (R_g^2 + X_g^2) + \sqrt{\left(1.5K_q X_g V_g \cos\delta + 1.5K_q R_g V_g \sin\delta - (R_g^2 + X_g^2)\right)^2 + 6K_q X_g(V_0 + K_q Q_{ref}) \cdot (R_g^2 + X_g^2)}}{2(R_g^2 + X_g^2)K_q X_g} \cdot$$

$$\left[R_g\left(\frac{1.5K_q X_g V_g \cos\delta + 1.5K_q R_g V_g \sin\delta - (R_g^2 + X_g^2) + \sqrt{\left(1.5K_q X_g V_g \cos\delta + 1.5K_q R_g V_g \sin\delta - (R_g^2 + X_g^2)\right)^2 + 6K_q X_g(V_0 + K_q Q_{ref}) \cdot (R_g^2 + X_g^2)}}{3K_q X_g}\right) - V_g \cos\delta\right] + X_g V_g \sin\delta$$

$$(9)$$

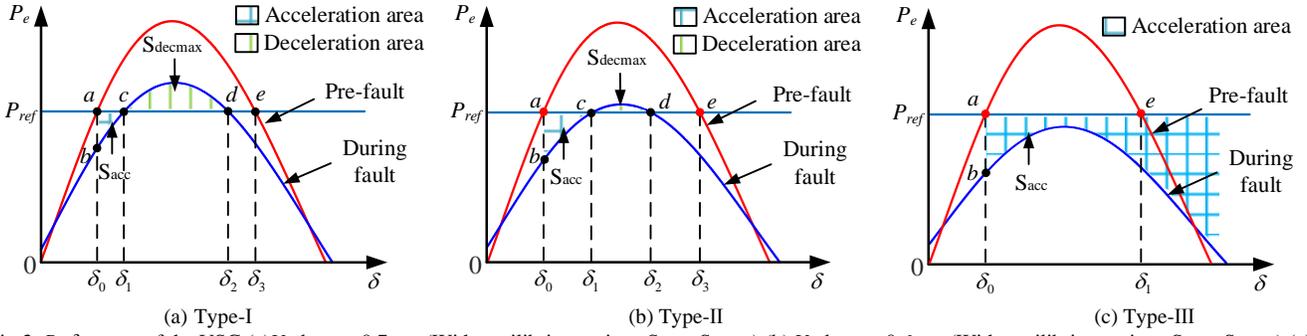

Fig.3. $P_e$-$\delta$ curves of the VSG (a) $V_g$ drop to 0.7p.u. (With equilibrium points, $S_{acc}$ < $S_{decmax}$) (b) $V_g$ drop to 0.6p.u. (With equilibrium points, $S_{acc}$ > $S_{decmax}$) (c) $V_g$ drop to 0.5p.u. (Without equilibrium points, no deceleration area exists.)

TABLE II PRIMARY PARAMETERS OF VSG

| parameter | value | parameter | value |
|---|---|---|---|
| Grid voltage $V_g$(rms)/V | 220 | Active power reference $P_{ref}$/kW | 100 |
| Fundamental frequency $f_o$/Hz | 50 | Reactive power reference $Q_{ref}$/var | 0 |
| Nominal voltage $V_o$/V | 311 | Filter inductance $L_f$/μH | 660 |
| Damping Coefficient $D$/ (N·m·s)/rad | 509.3 | Filter capacitor $C_f$/μF | 38.4 |
| Inertia Constant $H$ | 7.85 | Grid-side inductance $L_l$/mH | 3 |
| reactive droop coefficient $K_q$ | 0.0003 | Grid-side resistance $R_l$/Ω | 0.2 |

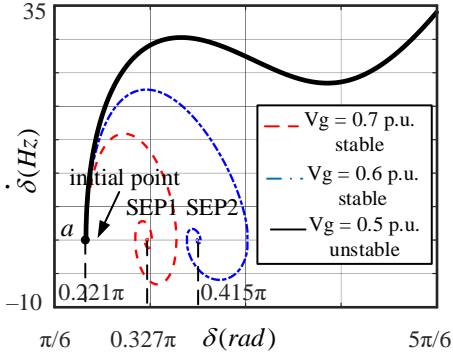

Fig.4. Phase portrait of VSG under different Vg dip fault where the dashed red line, the dashed blue line and the solid black line represent the phase trajectory when $V_g$ drop to 0.7p.u., 0.6p.u. and 0.5p.u, respectively

**Type-III** (Without equilibrium points): Fig.3(c) depicts the scenario where no equilibrium points exist after a large disturbance. In this case, since $P_{ref} > P_e$ always holds, the power angle $\delta$ will continuously increase due to the absence of deceleration area. Eventually, the VSG loses the synchronism with the grid.

[12] provides the fundamental principle and practical implementation approaches of the phase portrait. Therefore, when the parameters given in Table I are adopted, the phase trajectory of VSG during grid fault can be depicted in Fig.4. For Type-II transient stability problem, the phase portrait result reveals that VSG can maintain its stability while EAC prediction shows it will lose synchronization under the same parameter conditions. This is because, when using the EAC, the damping coefficient is not taken into consideration. In some scenarios, where VSG is with high damping, the application of the EAC for transient stability analysis results in inaccuracies, rendering an exact evaluation of its transient stability unattainable. Although the damping term and the inertia are considered in phase portrait method, it is difficult to explain the instability mechanism intuitively. In order to take these considerations into account, this paper uses TRM to estimate the DOA boundary of the grid-connected VSG, and analyzes the instability mechanism in detail combined with the phase trajectory. This approach provides a reliable analytical tool for assessing and judging the stability of VSG, which will be introduced in next Section.

## III. TRANSIENT STABILITY ANALYSIS WITH DOA ESTIMATION BASED ON TRAJECTORY REVERSING METHOD

### A. Trajectory Reversing Method for DOA Estimation

Recalling equation (3), the VSG is a typical second-order nonlinear system, which can be described as

$$\begin{cases} \dot{\delta} = \Delta\omega \\ \ddot{\delta} = \frac{1}{2H}\left(P_{ref} - P_e - D\dot{\delta}\right) \end{cases} \quad (10)$$

The backward integration of the VSG yields the following nonlinear dynamical system:

$$\begin{cases} \dot{\delta} = \Delta\omega \\ \ddot{\delta} = -\frac{1}{2H}\left(P_{ref} - P_e - D\dot{\delta}\right) \end{cases} \quad (11)$$

Equations (10) and (11) possess an identical topological structure, with the only difference being that the trajectories are in the opposite direction in the phase plane. For a second-order system where equilibrium points exist, the system may have SEPs as well as UEPs [23]. [24] employs TRM from a SEP, the result yields an enlarged attraction domain. Whereas, it cannot delineate the boundary. In order to delineate the boundary of DOA, in this paper, the method of employing TRM from the UEP is adopted. The specific implementation steps are as follows:

***step1:*** For a second-order VSG system, solve the equation (3) by setting the first derivative $\dot{\delta}$ and second derivative $\ddot{\delta}$ to zero respectively, i.e. $P_{ref} = P_e$. Equilibrium points ($\delta_0$,0) can be obtained only when the equation has a solution. Based on the characteristics of VSG, it is assumed that $V_{PCC}$ remains roughly constant. Thus, $\delta_0$ satisfies

$$P_{ref} = \frac{3}{2} \cdot \frac{V_{PCC}}{R_g^2 + X_g^2} \left[ R_g \left( V_{PCC} - V_g \cos\delta_0 \right) + X_g V_g \sin\delta_0 \right] \quad (12)$$

**step2:** Since the stability analysis of a nonlinear system is usually performed at the origin, the equilibrium point ($\delta_0$,0) should be shifted to the origin by setting $\delta_s = \delta - \delta_0$. Considering $\dot{\delta}_s = \dot{\delta} - 0$, $\ddot{\delta}_s = \ddot{\delta}$, the shifted VSG system can be described as

$$\ddot{\delta}_s = -\frac{D}{2H}\dot{\delta}_s + \frac{1}{2H}\left( P_{ref} - \frac{3}{2} \cdot \frac{V_{PCC}}{R_g^2 + X_g^2} \left[ R_g \left( V_{PCC} - V_g \cos(\delta_s + \delta_0) \right) + X_g V_g \sin(\delta_s + \delta_0) \right] \right) \quad (13)$$

Noticing that $\delta_s$ approaches 0, $\cos(\delta_s+\delta_0) \approx \cos\delta_0$, $\sin(\delta_s+\delta_0) \approx \delta_s \cdot \cos\delta_0 + \sin\delta_0$. (17) can be simplified as

$$\ddot{\delta}_s = -\frac{D}{2H}\dot{\delta}_s + \frac{1}{2H}\left( P_{ref} - \frac{3}{2} \cdot \frac{V_{PCC}}{R_g^2 + X_g^2} \left[ R_g \left( V_{PCC} - V_g \cos\delta_0 \right) + X_g V_g \sin\delta_0 + X_g V_g \delta_s \cos\delta_0 \right] \right) \quad (14)$$

By substituting (12) into (14), the shifted VSG system is given by

$$\ddot{\delta}_s = -\frac{D}{2H}\dot{\delta}_s - \frac{3V_{PCC}X_g V_g \cos\delta_0}{4H\left(R_g^2 + X_g^2\right)} \cdot \delta_s \quad (15)$$

Therefore, the Jacobian matrix of the shifted VSG system can be obtained, which yields

$$\begin{bmatrix} \dot{\delta}_s \\ \ddot{\delta}_s \end{bmatrix} = \begin{bmatrix} 0 & 1 \\ -\frac{3V_{PCC}X_g V_g \cos\delta_0}{4H\left(R_g^2 + X_g^2\right)} & -\frac{D}{2H} \end{bmatrix} \begin{bmatrix} \delta_s \\ \dot{\delta}_s \end{bmatrix} \quad (16)$$

The eigenvalues of the Jacobian matrix can be deduced, i.e.

$$\lambda_{1,2} = \frac{-D\sqrt{R_g^2 + X_g^2} \pm \sqrt{D^2\left(R_g^2 + X_g^2\right) - 12HV_{PCC}X_g V_g \cos\delta_0}}{4H\sqrt{R_g^2 + X_g^2}} \quad (17)$$

When $\cos\delta_0 < 0$, the equilibrium point has eigenvalues of both positive and negative real part, which indicates UEP. Conversely, when $\cos\delta_0 > 0$, the real part of the eigenvalues of this point is both negative with the following restriction

$$D^2\left(R_g^2 + X_g^2\right) \geq 12HV_{PCC}X_g V_g \cos\delta_0 \quad (18)$$

Such equilibrium point denotes SEP.

**step3:** Select several points uniformly in a small circular neighborhood around the UEP as the initial points. The amounts of initial points ought to be chosen carefully, as an insufficient number of these points could result in an inaccurate presentation of DOA boundary. 200 initial points are generally selected in this paper.

**step4:** Start from the selected initial points, integrate the equation (11) numerically and the nearly exact DOA boundary can be obtained.

Based on the parameters given in Table I, and the TRM steps for DOA boundary estimation, the results can be plotted as shown in Fig.5, where the solid blue line represents the estimated boundary.

As discussed in Section II, there are three types of transient stability problems when the grid fault occurs. The transient instability mechanism of VSG will be revealed in next section based on the estimated DOA boundary and the phase trajectory.

### B. Transient Instability Mechanism

Considering the critical event where the large signal disturbance occurs, causing different grid voltage drops. The DOA boundaries during the fault and after fault recovery can be plotted utilizing the aforementioned TRM steps, which are shown in Fig.6.

The blue solid line represents the post-fault DOA boundary and the black dashed line represents the DOA boundary after fault recovery, where the grid voltage restores to 1p.u. The phase trajectory of power angle $\delta$ during the fault can be plotted using the phase portrait method, which is presented by the solid black line in Fig.6. According to the position relationship between power angle phase trajectory and DOA before and after the disturbance, and whether DOA exists after the disturbance, the transient stability problems can be roughly divided into three types.

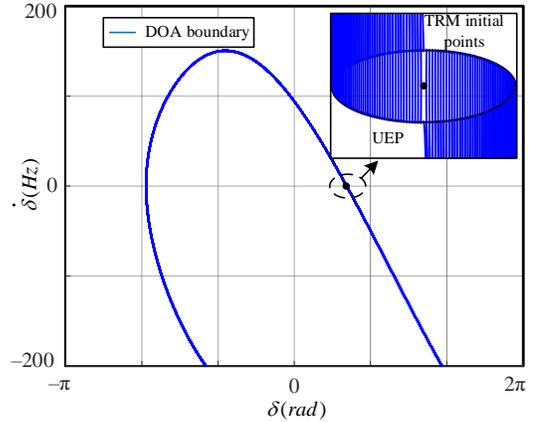

Fig. 5. DOA boundary estimation results with TRM.

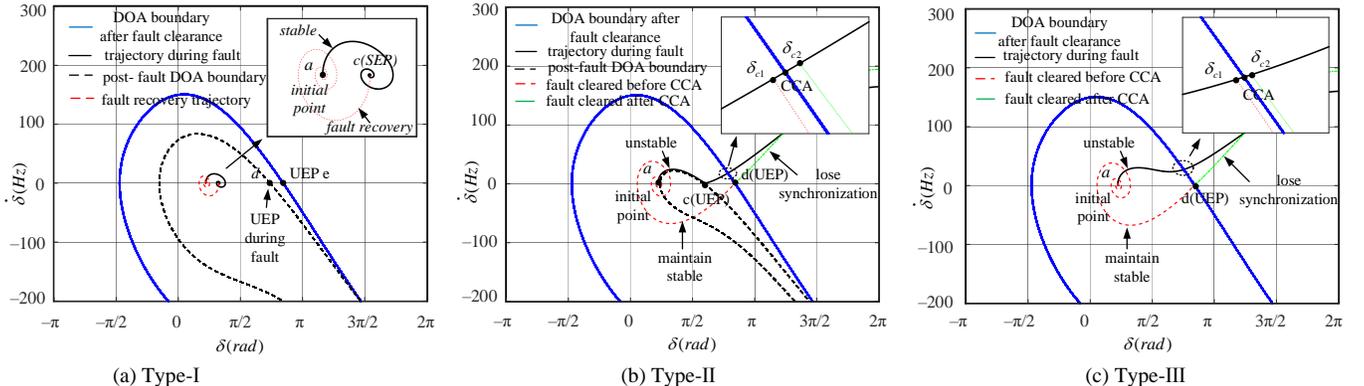

(a) Type-I    (b) Type-II    (c) Type-III

Fig. 6. Power angle phase portrait and DOA boundaries before and after grid fault. (a)The power angle trajectory is within the DOA range after disturbance. (b)The power angle trajectory is outside the DOA range after disturbance. (c)No DOA exists after disturbance.

*Type-I (the power angle trajectory is within the post-fault DOA range.):* When the grid voltage drops to 0.7p.u., as depicted in Fig.6(a), upon the advent of a large signal disturbance, the power angle, commencing from the initial SEP a, asymptotically stabilizes at the new SEP c. This convergence is facilitated by the fact that the power angle's trajectory during the fault remains securely within the post-fault DOA boundary. Consequently, the VSG is able to preserve transient stability in the presence of large disturbances in this case.

*Type-II (the power angle trajectory is outside the post-fault DOA range.):* As depicted in Fig.6(b), when the grid voltage drops to 0.57p.u., the power angle phase trajectory during the fault fails to converge to a new SEP. The VSG will lose synchronization with the power grid. This destabilization primarily occurs because the initial SEP and the power angle trajectory are located outside the post-fault DOA range. The intersection angle of the post-fault DOA boundary and the power angle phase trajectory is also identified as the critical clearing angle (CCA). $\delta_{c1}$ and $\delta_{c2}$ denote the fault clearing angle (FCA) where $\delta_{c1}$ < CCA and $\delta_{c2}$ > CCA. If the fault recovery is carried out at $\delta_{c1}$, the power angle phase trajectory lies outside the post-fault DOA range at this time, but it remains within the DOA range after grid voltage recovery. Therefore, the VSG can reconverge to the post-fault SEP c and recover transient stability after the occurrence of a large disturbance as depicted by the red dashed line in Fig.6(b). On the contrary, if FCA is larger than CCA, as exemplified by the green solid line in Fig.6(b), LOS will materialize on account of the power angle trajectory exceeding the DOA boundary. Thus, the mere existence of equilibrium points cannot guarantee the transient stability of the VSG, when the power angle phase trajectory lies outside the post-fault DOA range, prompt fault clearance is needed to avert potential system instability.

*Type-III (No DOA exists after disturbance.):* When the grid voltage drops to 0.5p.u., the VSG lacks equilibrium points, which implies the non-existence of DOA for VSG to converge upon. LOS will inevitably occur unless timely fault recovery measures are executed as shown in Fig.6(c). Likewise, whether VSG can maintain transient stability mainly depends on the magnitude of FCA and CCA. When the fault recovery is carried out at $\delta_{c1}$, where $\delta_{c1}$ < CCA, the VSG can maintain transient stability as depicted by the dashed red line in Fig.6(c). Conversely, when the fault recovery is carried out at $\delta_{c2}$, where $\delta_{c2}$ > CCA, the VSG will remain unstable.

### C. In-depth Analysis of the Key Parameters Influencing the DOA Range

The proposed TRM is also applicate to guide the design of VSG parameters. This section will investigate the impacts of key parameters on the DOA range and provide theoretical basis for the formulation of control strategy dedicated to the enhancement of transient stability. According to equation (9), the presence and positioning of the equilibrium points are independent of the equivalent damping coefficient $D$ and inertia constant $H$. However, they exert a discernible influence on the size of DOA in the company with the different voltage drop depth, the reactive droop coefficient $K_q$, the active power reference $P_{ref}$ and the $R_g/X_g$ ratio. Based on the aforementioned steps and the parameters given in Table I, the DOA boundaries with different system parameters can be plotted, as shown in Fig.7.

*1) Impact of damping coefficient D and inertia constant H*: In Fig.7(a), the DOA boundary of the VSG with different $D$ is depicted. As seen, an increment in $D$ augments the system's disturbance rejection capability, which is reflected in the expansion of the DOA boundary. While with the decreasing of $H$, the DOA boundary is observed to expand in Fig.7(b), leading to an enhancement in the transient stability performance of VSG. Besides, adjustments to $D$ and $H$ do not result in a displacement of the VSG's equilibrium points, hence, they retain an identical UEP.

*2) Impact of reactive droop coefficient $K_q$*: With the reduction of $K_q$, the DOA boundary is observed to expand, indicating an enhancement in the transient stability, as illustrated in Fig.7(c). According to equation (4), the variation in $K_q$ affects the amplitude of the output voltage of VSG. A diminished $K_q$ ensures that the fluctuations in $V_{PCC}$ are contained within a moderate range, and substantially bolster the transient stability of VSG, which is crucial for withstanding disturbances and averting the descent into instability.

*3) Impact of active power reference $P_{ref}$*: An increase of $P_{ref}$ results in the rightward shift of the SEP and the leftward shift of the UEP according to $P_e$-$\delta$ curves shown in Fig.3. This migration results in a gradual constriction of the DOA, culminating in the absence of a DOA when $P_{ref}$ and $P_e$ no longer intersect. Consequently, the system's transient stability performance will rapidly deteriorate as illustrated in Fig.7(d).

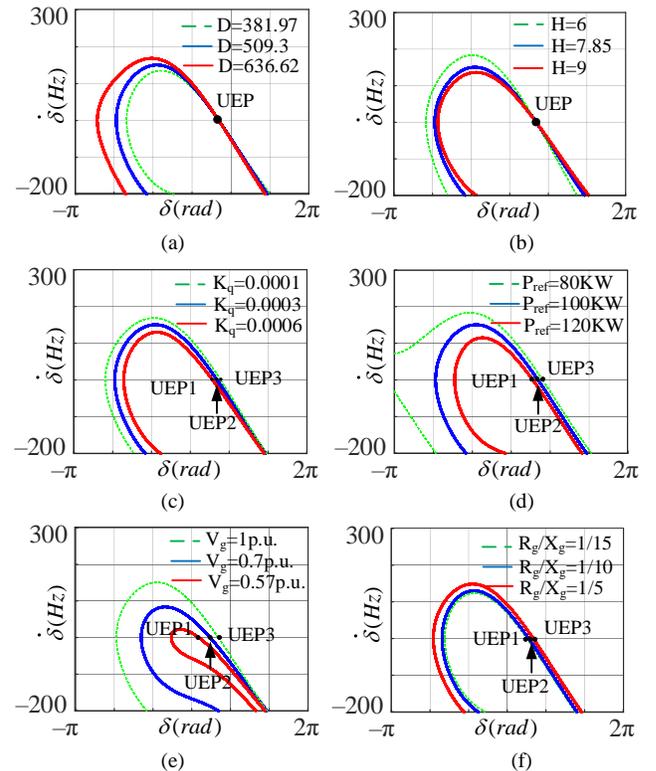

Fig. 7. DOA boundary with the variation of system parameters. (a) damping coefficient D. (b) inertia constant H. (c) reactive droop coefficient $K_q$. (d) Active power reference $P_{ref}$. (e) grid voltage drop depth. (f) $R_g/X_g$ ratio.

*4) Impact of grid voltage drop depth*: Fig.7(e) delineates the variation in the DOA boundaries in response to fluctuations in the grid voltage $V_g$. It is evident that as the severity of the voltage sag intensifies, $P_e$ decreases in accordance with equation (5). This results in a downward shift of the $P_e$ curve, causing the SEP to migrate to the right and the UEP to shift to the left. Consequently, as the SEP and UEP get closer, the DOA range diminishes progressively. This reduction continues until the SEP and UEP align or potentially nullify each other, culminating in the non-existence of DOA, reflecting a progressive deterioration in the system's transient stability.

*5) Impact of $R_g/X_g$ ratio*: According to equation (5), reduction in the $R_g/X_g$ ratio effectively decreases $P_e$, leading to a diminution of the DOA range as depicted in Fig.7(f). This will lead to a worse performance of the transient stability of VSG.

In summary, a larger damping coefficient $D$, a smaller inertia constant $H$, a smaller reactive power droop coefficient $K_q$, and appropriate values for the active power reference $P_{ref}$ and the $R_g/X_g$ ratio all contribute to enhancing the transient stability of VSG.

### D. A Comparative Study of CCA Results

The CCA is critical to the maintenance of system stability after a large disturbance. The determination of the CCA through EAC is predicated on the comparative analysis of the acceleration and deceleration areas. Focusing on the Type-II scenario, the VSG may exhibit transient instability in the presence of equilibrium points during a large disturbance. As depicted in Fig.8(a), in the event of a fault recovery at point $\delta_2$, where the grid voltage is restored to 1p.u., the FCA $\delta_2$ can be ascertained through the subsequent equation

$$\int_{\delta_0}^{\delta_1}(P_{ref}-P_e)d\delta - \int_{\delta_1}^{\delta_2}(P_e-P_{ref})d\delta - \int_{\delta_2}^{\delta_3}(P_e^*-P_e)d\delta \\ - \int_{\delta_3}^{\delta_4}(P_e^*-P_{ref})d\delta = 0 \quad (19)$$

where $\delta_0$, $\delta_1$, $\delta_3$ $\delta_4$ denote the angle corresponding to the intersection point of $P_e$ curve and $P_{ref}$ after disturbance and after fault recovery, respectively. $P_e^*$ represents $P_e$ curve after fault recovery. Equation (19) can be further written as

$$\int_{\delta_0}^{\delta_2}(P_{ref}-P_e)d\delta = \int_{\delta_2}^{\delta_3}(P_e^*-P_e)d\delta + \int_{\delta_3}^{\delta_4}(P_e^*-P_{ref})d\delta \quad (20)$$

When the parameters outlined in Table I are adopted, the CCA is calculated to approximate 1.784 rad. The DOA method facilitates a direct assessment of the CCA's magnitude through the intersection of the DOA boundary with the instability trajectory, yielding an approximate value of 2.461 rad.

In the context of the Type-III fault scenarios, equilibrium points are absent, as illustrated in Fig.8(b). When fault recovery is executed at point $\delta_1$, the CCA can be derived by

$$\int_{\delta_0}^{\delta_2}(P_{ref}-P_e)d\delta = \int_{\delta_2}^{\delta_3}(P_e^*-P_{ref})d\delta \quad (21)$$

where $\delta_0$, $\delta_2$ denote the angle corresponding to the intersection point of $P_e$ curve and $P_{ref}$ after disturbance and after fault recovery, respectively. $P_e^*$ represents $P_e$ curve after fault recovery. Similarly, the CCA can be deduced, which is approximately 1.69 rad. While utilizing the DOA method, the CCA is approximately 2.366 rad.

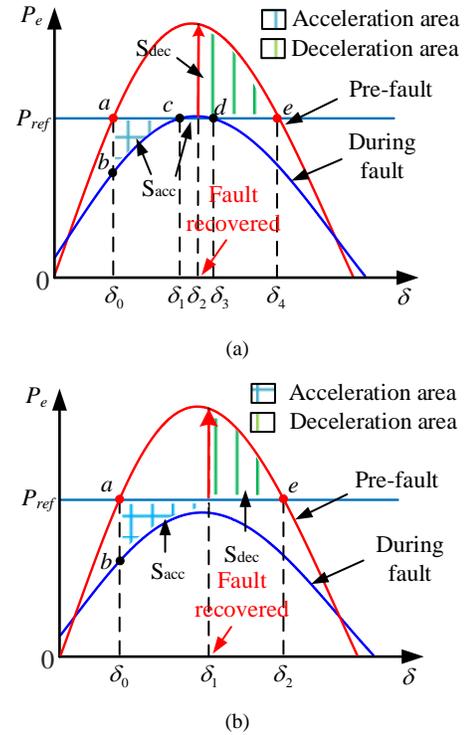

Fig. 8. $P_e$-$\delta$ curves of the VSG with fault clearance. (a) Vg drops to 0.57 p.u. (b) Vg drops to 0.5 p.u.

This comparison underscores the enhanced precision of the CCA as determined by the DOA method, which also circumvents the necessity for calculating the acceleration and deceleration areas, thereby diminishing the computational workload. Moreover, it boasts a high degree of intuitiveness and a reduced level of conservatism.

## IV. VERIFICATION

To validate the efficacy of the proposed analytical approach and the precision of the CCA results, a time-domain simulation model has been crafted within the MATLAB/SIMULINK environment, aligned with the system block diagram depicted in Fig.1 and the pertinent system parameters listed in Table I. The parameter $k_q$ is meticulously set to 0.0003, signifying that the voltage amplitude at the PCC fluctuates by a mere 0.03% in response to alterations in reactive power [25]. This calibrated selection is instrumental in ensuring the voltage at PCC point remains basically unchanged under large disturbances and that the DOA range of the system is as large as possible.

Fig.9 shows the simulation result for the Type-I stability problem where the system power angle phase trajectory is within the post-fault DOA boundary. When the grid voltage declines to 0.7p.u., the power angle can converge to the new SEP during the grid fault, which means that the VSG can maintain stability throughout the transient process, even in the absence of fault recovery.

Fig.10 and Fig.11 present the simulation results for the two distinct fault scenarios corresponding to Type-II and Type-III stability problems, respectively. When the grid voltage declines to 0.57p.u., the power angle trajectory lies outside the post-fault DOA range, leading to system instability. In order to maintain transient stability of the system during and after the grid faults, fault recovery must be carried out before CCA. As depicted in Fig.10(a), the grid voltage drops to 0.57p.u. at 1.5s and the grid voltage is restored to 1p.u. after fault recovery. When the FCA is less than the CCA derived from the DOA boundary and the phase trajectory during grid fault, indicating that the power angle trajectory has not yet exceeded the system's DOA boundary after fault recovery, the VSG can regain stability under this circumstance. Conversely, as depicted in Fig.10(b), if the FCA surpasses the CCA, the power angle trajectory exceeds the DOA boundary, and the VSG will persist in losing synchronization even if the grid voltage is restored to 1p.u..

Upon a further reduction of the grid voltage to 0.5p.u., due to the absence of equilibrium points after the disturbance, no new DOA boundary can be found, inevitably leading to transient instability during the disturbance. At this point, whether the system can maintain stability after fault recovery depends on whether fault recovery can be carried out before the power angle trajectory exceeds the initial DOA boundary. As depicted in Fig.11(a), when the FCA is less than the CCA, the VSG is capable of preserving transient stability after disturbance. Whereas, if the FCA surpasses the CCA, even with fault recovery, the VSG remains transiently unstable, which is presented in Fig11(b).

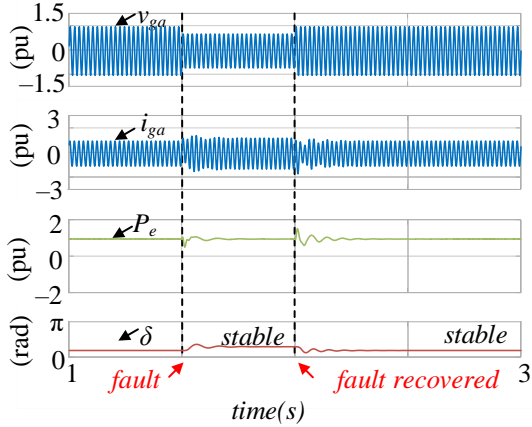

Fig. 9 Vg drop to 0.7p.u and system maintains stable.

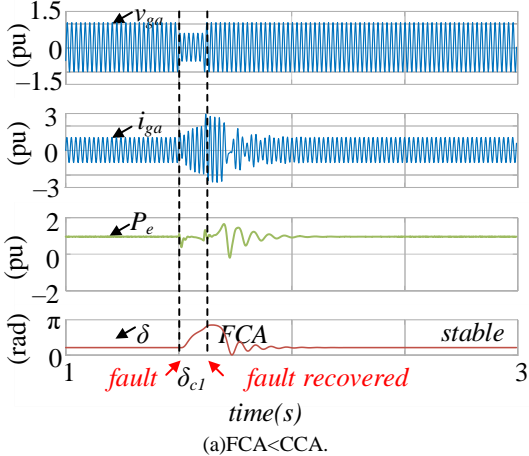

(a)FCA<CCA.

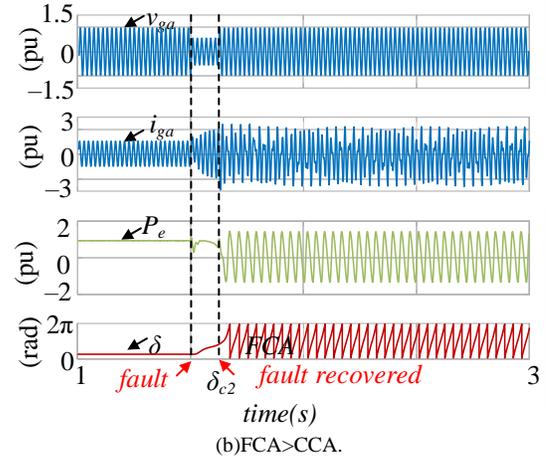

(b)FCA>CCA.

Fig. 10 Vg drop to 0.57p.u.(a) FCA<CCA and system maintain stable. (b) FCA>CCA. LOS.

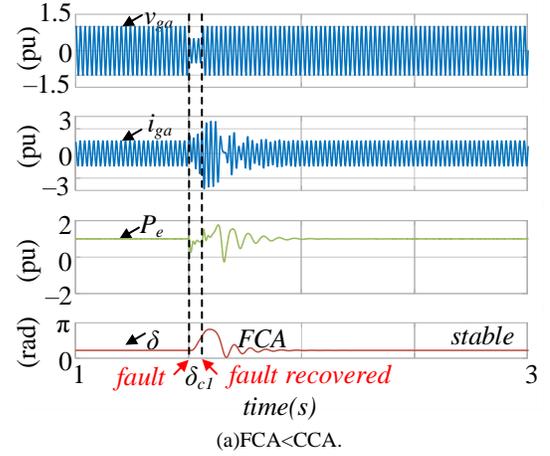

(a)FCA<CCA.

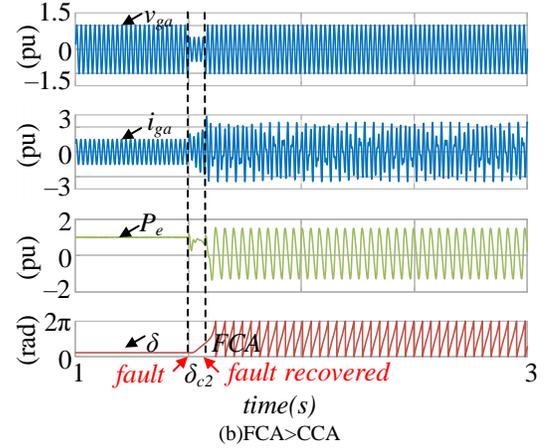

(b)FCA>CCA

Fig. 11 Vg drops to 0.5 p.u. (a)FCA<CCA and system maintain stable. (b)FCA>CCA LOS.

## V. CONCLUSION

In this paper, the different transient response types of VSG under voltage drop faults in the power grid have been studied. By employing the trajectory reversing method for domain of attraction estimation combined with the power angle phase trajectory during the grid fault, the reasons contributing to transient stability and transient instability are analyzed in detail. A simple approach to acquire the DOA of VSG is proposed, and the impact of different system parameter changes on DOA boundaries is analyzed through DOA in detail, which can offer promising insights for exploring methods to enhance transient stability. Furthermore, based on the DOA region estimated by TRM and the phase trajectories during large signal disturbance, the CCA result is acquired, which is more accurate than that obtained by EAC. Simulations have been presented to prove the accuracy of CCA estimated by the proposed method.

## References

[1] Wang X, Taul M G, Wu H, et al. Grid-synchronization stability of converter-based resources—An overview[J]. IEEE Open Journal of Industry Applications, 2020, 1: 115-134

[2] J. Quintero, V. Vittal, G. T. Heydt and H. Zhang, "The Impact of Increased Penetration of Converter Control-Based Generators on Power System Modes of Oscillation," in IEEE Trans. on Power Systems, vol. 29, no. 5, pp. 2248-2256, Sept. 2014.

[3] P. Ge, F. Xiao, C. Tu, Q. Guo, J. Gao and Y. Song, "Comprehensive transient stability enhancement control of a VSG considering power angle stability and fault current limitation," in CSEE Journal of Power and Energy Systems.

[4] P. Utkarsha, N. K. S. Naidu, B. Sivaprasad and K. A. Singh, "A Flexible Virtual Inertia and Damping Control Strategy for Virtual Synchronous Generator for Effective Utilization of Energy Storage," in IEEE Access, vol. 11, pp. 124068-124080, 2023.

[5] F. Wang, L. Zhang, X. Feng and H. Guo, "An Adaptive Control Strategy for Virtual Synchronous Generator," in IEEE Trans. on Industry Applications, vol. 54, no. 5, pp. 5124-5133, Sept.-Oct. 2018.

[6] X. Wang and F. Blaabjerg, "Harmonic Stability in Power Electronic-Based Power Systems: Concept, Modeling, and Analysis," in IEEE Trans. on Smart Grid, vol. 10, no. 3, pp. 2858-2870, May 2019.

[7] H. Wu and X. Wang, "Design-Oriented Transient Stability Analysis of PLL-Synchronized Voltage-Source Converters," in IEEE Trans. on Power Electronics, vol. 35, no. 4, pp. 3573-3589, April 2020.

[8] Z. Shuai, C. Shen, X. Liu, Z. Li and Z. J. Shen, "Transient Angle Stability of Virtual Synchronous Generators Using Lyapunov's Direct Method," in IEEE Trans. on Smart Grid, vol. 10, no. 4, pp. 4648-4661, July 2019.

[9] M. G. Taul, X. Wang, P. Davari and F. Blaabjerg, "An Overview of Assessment Methods for Synchronization Stability of Grid-Connected Converters Under Severe Symmetrical Grid Faults," in IEEE Trans. on Power Electronics, vol. 34, no. 10, pp. 9655-9670, Oct. 2019.

[10] Y. Tang, Z. Tian, X. Zha, X. Li, M. Huang and J. Sun, "An Improved Equal Area Criterion for Transient Stability Analysis of Converter-Based Microgrid Considering Nonlinear Damping Effect," in IEEE Trans. on Power Electronics, vol. 37, no. 9, pp. 11272-11284, Sept. 2022.

[11] C. Shen et al., "Transient Stability and Current Injection Design of Paralleled Current-Controlled VSCs and Virtual Synchronous Generators," in IEEE Trans. on Smart Grid, vol. 12, no. 2, pp. 1118-1134, March 2021.

[12] M. G. Taul, X. Wang, P. Davari and F. Blaabjerg, "An Efficient Reduced-Order Model for Studying Synchronization Stability of Grid-Following Converters during Grid Faults," 2019 20th Workshop on Control and Modeling for Power Electronics (COMPEL), Toronto, ON, Canada, 2019, pp. 1-7.

[13] X. Fu et al., "Large-Signal Stability of Grid-Forming and Grid-Following Controls in Voltage Source Converter: A Comparative Study," in IEEE Trans. on Power Electronics, vol. 36, no. 7, pp. 7832-7840, July 2021.

[14] D. Han, A. El-Guindy and M. Althoff, "Power systems transient stability analysis via optimal rational Lyapunov functions," 2016 IEEE Power and Energy Society General Meeting (PESGM), Boston, MA, USA, 2016, pp. 1-5.

[15] M. K. Bakhshizadeh, S. Ghosh, G. Yang and Ł. Kocewiak, "Transient Stability Analysis of Grid-Connected Converters in Wind Turbine Systems Based on Linear Lyapunov Function and Reverse-Time Trajectory," in Journal of Modern Power Systems and Clean Energy, vol. 12, no. 3, pp. 782-790, May 2024.

[16] N. G. Bretas and L. F. C. Alberto, "Lyapunov function for power systems with transfer conductances: extension of the Invariance principle," in IEEE Trans. on Power Systems, vol. 18, no. 2, pp. 769-777, May 2003.

[17] Zhang C, Molinas M, Li Z, et al. Synchronizing Stability Analysis and Region of Attraction Estimation of Grid-Feeding VSCs Using Sum-of-Squares Programming. Frontiers in Energy Research, 2020.

[18] L. Vetoshkin and Z. Müller, "Dynamic Stability Improvement of Power System by Means of STATCOM With Virtual Inertia," in IEEE Access, vol. 9, pp. 116105-116114, 2021.

[19] H. -D. Chiang, M. W. Hirsch and F. F. Wu, "Stability regions of nonlinear autonomous dynamical systems," in IEEE Trans. on Automatic Control, vol. 33, no. 1, pp. 16-27, Jan. 1988.

[20] A. Sheir, R. Milman and V. K. Sood, "Large Signal Stability of Grid-Tied Virtual Synchronous Generator Using Trajectory Reversing," 2022 IEEE Electrical Power and Energy Conference (EPEC), Victoria, BC, Canada, 2022, pp. 398-404.

[21] R. Genesio, M. Tartaglia and A. Vicino, "On the estimation of asymptotic stability regions: State of the art and new proposals," in IEEE Trans. on Automatic Control, vol. 30, no. 8, pp. 747-755, August 1985.

[22] X. Li et al., "The Largest Estimated Domain of Attraction and Its Applications for Transient Stability Analysis of PLL Synchronization in Weak-Grid-Connected VSCs," in IEEE Trans. on Power Systems, vol. 38, no. 5, pp. 4107-4121, Sept. 2023.

[23] Z. Dai, G. Li, M. Fan, J. Huang, Y. Yang and W. Hang, "Global Stability Analysis for Synchronous Reference Frame Phase-Locked Loops," in IEEE Trans. on Industrial Electronics, vol. 69, no. 10, pp. 10182-10191, Oct. 2022.

[24] R. Ma, J. Li, J. Kurths, S. Cheng and M. Zhan, "Generalized Swing Equation and Transient Synchronous Stability With PLL-Based VSC," in IEEE Trans. on Energy Conversion, vol. 37, no. 2, pp. 1428-1441, June 2022.

[25] H. Wu and X. Wang, "A Mode-Adaptive Power-Angle Control Method for Transient Stability Enhancement of Virtual Synchronous Generators," in IEEE Journal of Emerging and Selected Topics in Power Electronics, vol. 8, no. 2, pp. 1034-1049, June 2020.

[26] J. Rocabert, A. Luna, F. Blaabjerg and P. Rodríguez, "Control of Power Converters in AC Microgrids," in IEEE Trans. on Power Electronics, vol. 27, no. 11, pp. 4734-4749, Nov. 2012.

[27] Y. W. Li and C. -N. Kao, "An Accurate Power Control Strategy for Power-Electronics-Interfaced Distributed Generation Units Operating in a Low-Voltage Multibus Microgrid," in IEEE Trans. on Power Electronics, vol. 24, no. 12, pp. 2977-2988, Dec. 2009.

[28] R. Genesio and A. Vicino, "New techniques for constructing asymptotic stability regions for nonlinear systems," in IEEE Trans. on Circuits and Systems, vol. 31, no. 6, pp. 574-581, June 1984.

[29] P. Ge, C. Tu, F. Xiao, Q. Guo and J. Gao, "Design-Oriented Analysis and Transient Stability Enhancement Control for a Virtual Synchronous Generator," in IEEE Trans. on Industrial Electronics, vol. 70, no. 3, pp. 2675-2684, March 2023.